\documentclass[useAMS,usenatbib]{mn2e}

\usepackage{graphics}
\usepackage{epsfig}

\title[Dynamical Masses]{The AIMSS Project II:
Dynamical-to-Stellar Mass Ratios Across the Star Cluster - Galaxy Divide}
\author[D. A. Forbes et al.]{Duncan A. Forbes$^{1}$\thanks{E-mail:
dforbes@swin.edu.au}, Mark A. Norris$^{2}$, Jay Strader$^{3}$, Aaron
J. Romanowsky$^{4,5}$, 
\newauthor
Vincenzo Pota$^{1,5}$, 
Sheila J. Kannappan$^{6}$, Jean P. Brodie$^{5}$, Avon Huxor$^{7}$ 
\\
$^{1}$Centre for Astrophysics \& Supercomputing, Swinburne University, Hawthorn VIC 3122, Australia\\
$^{2}$Max Planck Institut fur Astronomie, Konigstuhl 17, D-69117,
Heidelberg, Germany\\
$^{3}$Department of Physics and Astronomy, Michigan State
University, East Lansing, Michigan 48824, USA\\
$^{4}$Department of Physics and Astronomy, San Jos\'e State
University, One Washington Square, San Jose, CA 95192, USA\\
$^{5}$University of California Observatories, 1156 High Street,
Santa Cruz, CA 95064, USA\\
$^{6}$Department of Physics and Astronomy, UNC-Chapel Hill,
CB3255, Phillips Hall, Chapel Hill, NC 27599, USA\\
$^{7}$Astronomisches Rechen-Institut, Zentrum fur Astronomie der
Universitat Heidelberg, Monchstrasse 12-14, D-69120 Heidelberg, Germany\\
}
\begin{document}


\pagerange{\pageref{firstpage}--\pageref{lastpage}} \pubyear{2002}

\maketitle

\label{firstpage}

\begin{abstract}

The previously clear division between small galaxies and
massive star clusters is now occupied by objects called ultra
compact dwarfs (UCDs) and compact ellipticals (cEs). Here we
combine a sample of UCDs and cEs with velocity dispersions from
the AIMSS project with literature data to explore their
dynamical-to-stellar mass ratios. 

We confirm that the mass ratios of many UCDs in the stellar mass range 
10$^6$ -- 10$^9$ M$_{\odot}$ are
systematically higher than those for globular clusters which have
mass ratios near unity. 
However, at the very highest masses in our sample,
i.e. 10$^9$ -- 10$^{10}$ M$_{\odot}$, we find that cE galaxies
also have mass
ratios of close to unity, indicating their central regions 
are mostly composed of
stars. 

Suggested explanations for the 
elevated mass ratios of UCDs 
have included a variable
IMF, a central black hole, and the presence of dark
matter.  
Here we present another possible explanation, i.e. tidal stripping. 
Under various assumptions, we find that the apparent variation in
the mass ratio with stellar
mass and stellar density can be qualitatively reproduced by published
tidal stripping simulations of a dwarf elliptical galaxy. In the
early stages of the stripping process the galaxy is unlikely to
be in virial equilibrium. 
At late stages, 
the final remnant resembles the properties of $\sim$10$^7$ M$_{\odot}$ UCDs.
Finally, we discuss the need for more detailed realistic
modelling of tidal stripping over a wider range of parameter
space, and observations to further test the stripping
hypothesis.

\end{abstract}


\section{Introduction}

The rate of discovery of new types of low mass stellar systems
over the last 15 years has been remarkable. These new systems 
include Extended Clusters (ECs), Faint Fuzzies (FFs), Diffuse
Star Clusters (DSCs) and Ultra
Compact Dwarfs (UCDs). The latter have sizes and/or masses that approach
those of dwarf ellipticals (dEs) and compact ellipticals (cEs).
This discovery process usually begins with imaging that
identifies candidates with {\it inferred} properties of size and
luminosity that occupy a previously empty, or sparse, region of
size-luminosity parameter space. The Hubble Space Telescope
(HST), with its ability to partially resolve objects of size
greater than 3 pc out to 40 Mpc distance, has played a key role.
The next step is spectroscopic confirmation that the object is
indeed associated with a larger host galaxy/group and is not
merely a background object 
seen in projection.  The scaling relation between physical
size and luminosity of confirmed objects can then be examined,
with the caveat that selection bias needs to be understood.  

A further fundamental parameter of galaxies and star clusters alike
is their internal velocity dispersion. Measuring this usually requires
dedicated spectroscopic follow-up with a high resolution
spectrograph and long exposure times. As well as probing the
velocity dispersion-luminosity scaling relation (i.e. extending
the Faber-Jackson relation into the low mass regime), one can
calculate dynamical mass and contrast it with stellar mass
estimates to gain insight on issues such as the dark matter
content and/or the stellar Initial Mass Function (IMF).  
After a few individual objects have been
studied in this way, larger, and statistically-complete, samples
of objects can eventually be examined.

As recently as 2007/8, the number of known cEs was only half a
dozen (Chilingarian et al. 2007) and UCDs numbered around two
dozen (Mieske et al. 2008; Dabringhausen et al. 2008; Forbes et
al. 2008) with a largely empty gap in the parameter space of
size-luminosity-velocity dispersion between them. This gap 
has been filled over the
years (e.g. Chiboucas et al. 2011; Brodie et al. 2011; Forbes et
al. 2013), but velocity dispersions for objects within the
gap have not kept pace as they tend to be observed individually
or in small numbers (Chilingarian \& Mamon 2008; Price et al.
2009; Forbes et al. 2011; Penny et al. 2014).

Most recently, Norris et al. (2014; N14) has identified a number of 
UCD and cE candidates from HST archive imaging with a 
variety of telescopes used to confirm their distances.
Velocity dispersions have been measured for over two dozen of them. 
Combining with the literature, a large number of
confirmed objects now exist which fill the gap in velocity
dispersion-luminosity parameter space between previously known
UCDs and cEs. 

In N14 we found many new objects to have lower
stellar masses for a given velocity dispersion compared to dEs
and normal elliptical galaxies. 
This is qualitatively consistent with UCDs and cEs being
the tidally-stripped remnants of dEs and ellipticals, respectively (Bekki
et al. 2003; Goerdt et al. 2008; Pfeffer \& Baumgardt 2013; Faber
1973; Bender, Burstein \& Faber 1992; Huxor et al. 2011).  For example, the
simulations of Pfeffer \& Baumgardt (2013) placed a nucleated
dE galaxy on various orbits in a Virgo cluster-like potential.
After tidal stripping of the stars (there is no dark matter in
their model) the final remnants have sizes and
luminosities  similar
to those of observed UCDs and in some cases the remnants are so
small that they resemble bright GCs (i.e. size $\sim$ 5 pc and M$_V$
$\sim$ --9.5). Unfortunately they did not predict velocity
dispersions. The objects discovered by N14 came from a range of
environments including the field which is dominated by late type
galaxies. Thus the progenitors of UCDs likely include disk
galaxies with nuclei and/or bulges. 

In N14 we also found evidence to support the idea of a gradual
transition from in-situ formed old star clusters (globular
clusters) to free-floating remnant nuclei or bulges, as one
probes masses from 10$^4$ M$_{\odot}$ to 10$^{10}$
M$_{\odot}$. UCDs in the mass range 
10$^6$-- 7 $\times$ 10$^7$ M$_{\odot}$ include both
origins, but for masses $> 7 \times 10^7$
M$_{\odot}$ they are predominately remnants of stripped
galaxies. We note that the transition is not only one of mass 
but also of size (Brodie et al. 2011; Forbes et al. 2013). 


Previous observations of UCDs suggest that they have elevated
dynamical-to-stellar mass ratios, and the ratio increases with
object mass (e.g. Mieske et al. 2013 and references therein).
Possible explanations, which may all be a consequence of an
origin in tidal stripping,  
include the presence of central black holes, 
stars with a non-universal IMF and dark matter. 

Central massive black holes are now known to be a common
occurrence in large galaxies.  
Recently, Mieske et al. (2013) concluded that central black holes
with a mass some 10-15 per cent of the current UCD mass could 
explain the elevated ratios. This is consistent with UCDs being
the stripped remnants of $\sim$10$^9$ M$_{\odot}$ galaxy progenitors, based on the black 
hole--galaxy scaling relation.

The possibility of a different IMF in UCDs compared to globular
clusters, with either bottom-heavy or top-heavy variants, has been suggested 
(Mieske et al. 2008). 
Recent support for a bottom-heavy IMF (more low mass stars) comes
indirectly from observations of the cores of giant ellipticals
(e.g. van Dokkum \& Conroy 2010). In favour of a top-heavy IMF (more stellar remnants), 
Dabringhausen et al. (2012) recently showed that UCDs have low
mass X-ray binary rates up to ten times those expected for a globular cluster-like
IMF (but see also Phillips et al. 2013 for an alternative conclusion). 

If tidal stripping has removed
most of the initial galaxy's mass then the remnant will have 
a stellar population similar to that of the progenitor galaxy core.  
If this is the case, then the assumption of
the same IMF for all UCDs gives a relatively lower stellar mass for
the higher luminosity UCDs, thus effectively raising their apparent 
dynamical-to-stellar mass ratio. 

Murray (2009) and Tollerud et al. (2011) have 
argued that if UCDs formed with a NFW-like dark
matter profile (Navarro, Frenk \& White 1997) 
then the expected dark matter density in the
central regions would be some hundred times less than that
observed for the stellar density, making dark matter relatively
unimportant and unlikely to be the cause of the elevated ratios.
Furthermore, Baumgardt \& Mieske (2008) showed that any dark matter can be effectively 
`pushed out' of a globular cluster even if it was present at formation.  
However, if UCDs formed from the tidal stripping of a larger galaxy, then
the remnant UCD may still contain some of the progenitor's dark
matter.
For example, in the simulations by Goerdt et
al. (2008) of the tidal stripping of a nucleated dwarf disk
galaxy, gas that falls into the remnant core 
can effectively drag dark matter from larger radii into the
core region. In the case of a gas-free progenitor dwarf
elliptical galaxy the central regions of the remnant UCD
would be expected to remain relatively dark matter free (Forbes et al. 2011).  
The search for dark matter in UCDs has the best chance of success
in those UCDs with {\it low} stellar densities, i.e. low luminosity
UCDs with large sizes (Willman \& Strader 2012).


\begin{table*}
\caption{Physical properties of AIMSS compact stellar systems
from Norris et al. (2014).}
\begin{tabular}{lcccccc}
\hline
ID & M$_V$ & R$_e$ & $\sigma_0$ & M$_{dyn}$ & M$_{\ast}$ & Ratio\\ 
 & (mag) &  (pc) & (km/s) & (M$_{\odot})$ & (M$_{\odot})$ & \\
\hline
                          NGC~0524-AIMSS1  &    -12.6  &  39.9$\pm$3.8  &   36.3$\pm$2.9 &   7.95$\pm$2.0 $\times$10$^7$ &  4.96$\pm$0.11 $\times$10$^7$  &     1.60$\pm$0.44\\           
                          NGC~0703-AIMSS1  &    -15.0  & 164.7$\pm$28.4  &   20.1$\pm$7.0 &  1.01$\pm$0.87 $\times$10$^8$  & 3.13$\pm$0.98 $\times$10$^8$   &   0.32$\pm$0.38\\            
                          NGC~0741-AIMSS1  &    -17.6  & 311.7$\pm$55.0  &   79.8$\pm$4.4 &  2.99$\pm$0.86 $\times$10$^9$  & 5.96$\pm$0.40 $\times$10$^9$   &   0.50$\pm$0.18\\          
                          NGC~1128-AIMSS1  &    -15.6  &  76.0$\pm$10.9  &   76.9$\pm$7.0 &  6.79$\pm$2.2 $\times$10$^8$  & 7.50$\pm$0.16  $\times$10$^8$   &   0.90$\pm$0.31\\        
			  NGC~1128-AIMSS2  &    -17.9  &  484.8$\pm$69.2  &  54.0$\pm$3.0 &   2.14$\pm$0.54 $\times$10$^9$ &  4.73$\pm$0.75 $\times$10$^9$  &    0.45$\pm$0.19\\
                            NGC~1132-UCD1  &    -14.8  &  84.3$\pm$12.1  &   93.5$\pm$9.5 &  1.11$\pm$0.39 $\times$10$^9$  & 3.28$\pm$0.88 $\times$10$^8$   &    3.40$\pm$2.09\\              
                          NGC~1172-AIMSS1  &    -11.6  &   6.4$\pm$0.6  &   52.4$\pm$14.0 & 2.66$\pm$1.67 $\times$10$^7$  &  6.84$\pm$3.11 $\times$10$^6$   &    3.88$\pm$4.20\\ 
                          Perseus-UCD13  &    -12.8  &  88.6$\pm$8.6  &   38.0$\pm$9.0  &  1.93$\pm$1.10 $\times$10$^8$  & 2.72$\pm$1.10 $\times$10$^7$   &    7.10$\pm$6.92\\                   
                          NGC~2768-AIMSS1  &    -12.1  &   6.4$\pm$0.7  &   45.3$\pm$5.5 &  1.99$\pm$0.70 $\times$10$^7$  & 5.43$\pm$3.00 $\times$10$^6$   &    3.65$\pm$3.30\\              
                          NGC~2832-AIMSS1  &    -14.9  &  46.4$\pm$6.7  & 133.7$\pm$13.2 & 1.25$\pm$0.43 $\times$10$^9$ &  2.37$\pm$0.86 $\times$10$^8$  &     5.28$\pm$3.72\\                
                          NGC~3115-AIMSS1  &    -11.3  &   8.6$\pm$0.4  &   41.6$\pm$2.1 &  2.25$\pm$0.33 $\times$10$^7$  & 1.09$\pm$0.33 $\times$10$^7$   &    2.07$\pm$0.93\\              
                            NGC~3923-UCD1  &    -12.4  &  12.3$\pm$0.3  &   43.4$\pm$2.8 &  3.50$\pm$0.54 $\times$10$^7$  & 1.97$\pm$0.56 $\times$10$^7$   &    1.77$\pm$0.77\\              
                            NGC~3923-UCD2  &    -11.9  &  13.0$\pm$0.2  &   26.9$\pm$4.2 &  1.42$\pm$0.47 $\times$10$^7$  & 6.53$\pm$2.15 $\times$10$^6$   &    2.17$\pm$1.42\\              
                            NGC~3923-UCD3  &    -11.3  &  14.1$\pm$0.2  &   19.0$\pm$4.4 &  7.69$\pm$3.67 $\times$10$^6$  & 2.37$\pm$0.80 $\times$10$^6$   &    3.24$\pm$2.64\\              
                          NGC~4350-AIMSS1  &    -12.2  &  15.4$\pm$0.1  &   25.4$\pm$9.0 &  1.50$\pm$1.07 $\times$10$^7$  & 1.57$\pm$0.57 $\times$10$^7$   &    0.95$\pm$1.02\\              
                            NGC~4546-UCD1  &    -12.9  &  25.5$\pm$1.3  &   20.0$\pm$2.3 &  1.54$\pm$0.43 $\times$10$^7$  &  3.59$\pm$0.86 $\times$10$^7$   &   0.42$\pm$0.21\\              
                          NGC~4565-AIMSS1  &    -12.4  &  17.4$\pm$1.4  &   15.3$\pm$9.0 &  6.16$\pm$7.74 $\times$10$^6$  & 8.19$\pm$0.31  $\times$10$^6$   &   0.75$\pm$0.97\\              
			  NGC~4621-AIMSS1  &    -11.9  &  10.2$\pm$0.4 &   41.9$\pm$5.4  &   2.71$\pm$0.80 $\times$10$^7$ &  1.64$\pm$0.38 $\times$10$^7$  &     1.65$\pm$0.87\\
                          M60-UCD1        &    -14.2  &  27.2$\pm$1.0  &   63.9$\pm$1.9 & 1.68$\pm$0.16 $\times$10$^8$   & 1.80$\pm$0.21 $\times$10$^8$	&0.93$\pm$0.19\\                
                          NGC~7014-AIMSS1  &    -15.2  & 329.8$\pm$23.6  &	19.0$\pm$5.8 & 1.80$\pm$1.22 $\times$10$^8$  & 2.99$\pm$0.98 $\times$10$^8$    &  0.60$\pm$0.60\\
\hline                         
                               UCD3/F-19  &  -13.5  &  86.5$\pm$6.2  &   26.6$\pm$4.9  &     9.25$\pm$4.07 $\times$10$^7$	&4.96$\pm$1.15 $\times$10$^7$	&1.87$\pm$1.25\\   
			  NGC~2832-cE	  &    -17.8   & 375.3$\pm$54.4 & 97.4$\pm$2.9	& 5.38$\pm$1.10 $\times$10$^9$   &2.27$\pm$0.51 $\times$10$^9$       &2.37$\pm$1.01\\
			  NGC~2892-AIMSS1  &    -18.9   & 580.9$\pm$85.0	 &137.5$\pm$3.7	&1.66$\pm$0.33 $\times$10$^{10}$   &1.09$\pm$0.12 $\times$10$^{10}$        &1.53$\pm$0.47\\
			  NGC~3268-cE1	  &   -15.9    & 299.9$\pm$21.9	 &33.3$\pm$13.6	&5.03$\pm$4.47 $\times$10$^8$  &1.30$\pm$0.26 $\times$10$^8$        &3.86$\pm$4.20\\
                           Sombrero-UCD1   &  -12.3 &  14.7$\pm$1.4  &   39.5$\pm$3.6  	&3.47$\pm$0.96 $\times$10$^7$	&1.64$\pm$0.41 $\times$10$^7$	&2.11$\pm$1.11\\
                                   M59cO   &  -13.4 &   35.2$\pm$1.2  &   25.7$\pm$2.2 	&3.19$\pm$0.66 $\times$10$^7$	&7.49$\pm$0.11 $\times$10$^7$	&0.42$\pm$0.09\\
 			  ESO383-G076-AIMSS2 & -17.4   & 652.2$\pm$57.5	 &97.5$\pm$8.9	&9.37$\pm$2.53 $\times$10$^9$   &2.60$\pm$0.53 $\times$10$^9$       &3.59$\pm$1.70\\
\hline
\end{tabular}
\\
Notes: Object name, V-band magnitude, stellar mass, 
effective radius and average uncertainties are from Norris et al. (2014). For central
velocity dispersion see Section 2.1, and for dynamical mass see
Section 4. Ratio is dynamical-to-stellar mass ratio, with the error
calculated from the measurement uncertainties in velocity dispersion, 
effective radius and stellar mass. Objects in the lower part of the table 
are those that have been re-observed or are serendipitous.
\end{table*}

Mieske et al. (2013)  recently carried out a study of UCDs in the 
Centaurus A group, and Fornax and Virgo clusters.
In addition to reproducing the trend of a rising
mass ratio with object mass, they 
claimed a bimodal structure in the
mass ratio of the lower mass ($<$ 10$^7$ M$_{\odot}$) UCDs. They suggested
that this was consistent with the idea that low mass UCDs were a
combination of stripped dwarf galaxies (with high mass ratios) and massive GCs 
(with low mass ratios). In this picture, only the high mass ($>$
10$^7$ M$_{\odot}$) UCDs
reveal elevated ratios consistent with a pure stripping origin. 
With the enlarged UCD and cE dataset of N14, we will re-test the 
claims for a rising trend and bimodality in the mass ratios of
UCDs. We also will address two key outstanding
issues relating to UCDs and cEs, namely 1) what is their origin? 
and 2) what is the cause of their elevated dynamical-to-stellar mass
ratios? 

In the next
two sections we describe the new AIMSS data and additional
data from the literaure. Section 4 describes our dynamical mass
calculation. Section 5 discusses UCD formation via tidal stripping before
presenting our results in Section 6. Our conclusions and thoughts on future
work are given in Section 7. 


\section{The AIMSS sample of compact stellar systems}

The Archive of Intermediate Mass Stellar Systems (AIMSS) targets
compact stellar system candidates identified in the HST archive
for spectroscopic followup (Norris \& Kannappan 2011). The
candidates are selected to have an inferred M$_V$ $<$ --10 and
effective half light radius R$_e$ that is twice the HST
resolution limit of 0.1$^{''}$. Objects must also be relatively
round ($\epsilon$ $<$ 0.25) and lie within 150 kpc in projection
of a large galaxy (M$_V$ $<$ --15). No colour selection is
applied. Objects with apparent magnitudes V $<$ 21.5 are
targetted for spectroscopic follow-up. Aperture velocity dispersion
measurements are available for 27 objects in N14;  
20 are new AIMSS objects and 7 are 
re-observed or serendipitous objects. 
For each object we list the V-band magnitude, total stellar mass
and effective radius from N14 in Table 1. 
To calculate stellar mass N14 used the code of Kannappan et al. (2013). 
This code uses the optical and infrared 
magnitudes, combined with a grid of stellar population models
from Bruzual \& Charlot (2003) covering ages from 5 My to 13.5
Gyr, and metallicities Z from 0.008 to 0.05 assuming a diet Salpeter
IMF (which is similar to a Kroupa IMF for the purposes of calculating 
stellar masses). 
Had a normal, rather than diet, Salpeter IMF been adopted, the 
stellar masses would be systematically higher by $\sim$0.15 dex (and 
dynamical-to-stellar mass ratios lower).
The effective radii 
come from fits to HST images of each object
as determined by N14. 

\subsection{Calculating central velocity dispersions}

The new velocity dispersion measurements presented in N14
are `raw' in the sense that they are the value
measured within the slit aperture quoted. The angular size of the
objects of interest are on the same order as the slit, which is
also comparable to our typical seeing. Thus in order to derive
central velocity dispersions, we need to correct for light
loss and the intrinsic surface brightness profile of the
object. We use the aperture size, seeing and half light radius
quoted in N14 and a King profile to make the corrections following 
method of Strader et al. (2011).  
When a Sersic profile provides the best fit, we define the 
central velocity dispersion as the integrated value within R$_e$/10.   
Although these 
corrections are typically a few percent, the final dynamical
mass depends on the corrected velocity dispersion squared.
Our final central velocity dispersions and  
measurement uncertainties for the 27 AIMSS objects are included in Table 1. 

\section{Literature Data}

As well as the objects listed in Table 1, we include various
datasets from the literature. 

In the Appendix Table A1, we list UCDs from the compilation of 
Mieske et al. (2013). 
They list {\it aperture} velocity
dispersions in their table 3 for the bulk of these objects. 
In Table A1 we list the {\it central}
velocity dispersions kindly supplied by Mieske \& Baumgardt
(2014, priv. comm.), supplemented by a few central values 
from Taylor et al. (2010). The errors quoted are measurement
uncertainties only. The V-band magnitude, stellar mass and effective radius
come from N14. 

We exclude half a dozen Cen A objects from the original Mieske et
al. (2013) list as they have M$_V$ $>$ --10. Although these objects
could in principle be included in our analysis 
as GCs, we prefer to keep the GC sample 
homogeneous using only GCs from either the Milky Way or M31. 

For other objects 
we take their properties, including their 
`central' velocity dispersions, from N14.
Briefly, we use Misgeld \& Hilker (2011) and
ATLAS$^{\rm 3D}$ (Cappellari et al. 2011) 
for giant early-type galaxies, with dwarf ellipticals coming 
from Geha et al. (2002, 2003), Toloba et al. (2012) and Forbes et
al. (2011). Local Group dwarf data are taken from Tollerud et
al. (2013 and references therein). Compact ellipticals come from
a variety of sources with the majority from Chilingarian et
al. (2009) and Price
et al. (2009). Milky Way
GCs come from the compilation of Harris (2010) and 
M31 GCs from Strader et al. (2011). We note that N14 assumed all GCs to 
be uniformly old, which for a large sample of M31 and Milky Way GCs is a 
reasonable assumption.  
We do not consider young massive clusters (which are not old
stellar populations) nor
nuclear star clusters (for which very few velocity dispersions exist) 
in this work.

The size-stellar mass distribution for our entire sample is shown
in Figure 1. Here we only show objects for which we have central 
velocity dispersions; for larger samples of objects with sizes
and masses (or luminosities), but which lack velocity dispersions, see
figure 13 of N14 (also Forbes et al. 2008; Brodie et al. 2011; 
Bruens \& Kroupa 2012). Figure 1 shows objects traditionally
classified as dwarf spheroidal (dSph), dwarf elliptical (dE),
giant elliptical (gE), globular cluster (GC), ultra compact dwarf
(UCD) and compact elliptical (cE), although the definition of
such objects is often poorly defined and they can overlap in
size-mass parameter space. For dSph, dE and gE galaxies we simply
use their classification as defined by the literature sources
listed above. Here we consider a GC to have stellar mass below 
10$^6$ M$_{\odot}$. Such objects also have sizes R$_e$ $<$ 10
pc (a notable exception being the Milky Way GC NGC 2419 which is  
considered by some to be a UCD).
We consider cEs to have stellar masses $>$ 10$^9$
M$_{\odot}$ and R$_e$ $<$ 800 pc. UCDs are taken to have
masses intermediate between those of GCs and cEs, irrespective of
their size (however we also the include the size of a UCD
in our analysis below, e.g. many low mass UCDs have R$_e$ $<$ 10
pc). The Figure also shows the Zone of Avoidance, i.e. a high
density region of parameter space that is so far largely unoccupied 
(see discussion in N14).

\begin{figure}
\begin{center}
\includegraphics[scale=0.33,angle=-90]{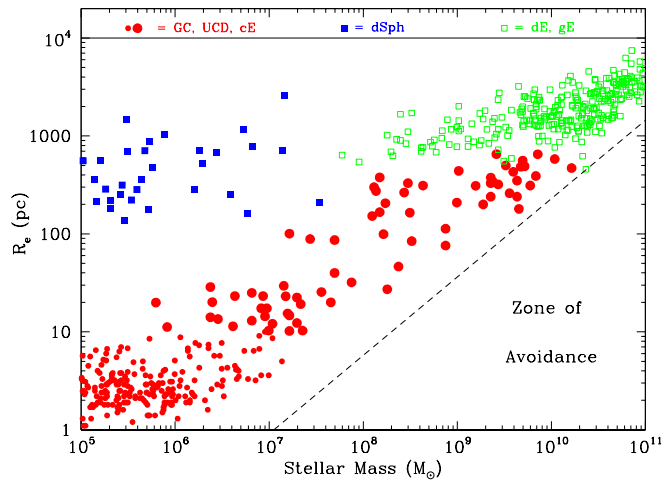}
\caption{Size vs stellar mass for the entire sample of 
old, pressure supported systems with available central velocity dispersions. 
Blue squares denote Local Group dSph galaxies, and green open squares
denote dE and gE galaxies.   
Red circles denote GCs, UCDs and cEs, 
with larger circles for objects with R$_e$ $>$ 10 pc. 
The diagonal dashed line demarks the constant surface density 
edge of the Zone of Avoidance, i.e. a high density region of
parameter space that appears to be unoccupied.  
There is some overlap in size-mass parameter space between the
traditional classification of different types of object.
}
\end{center}
\end{figure}

\section{Dynamical Masses}

Dynamical mass estimates for pressure-dominated systems
can be obtained using the expression:
\begin{equation}
M_{\rm dyn} = C G^{-1} \sigma^2 R,
\end{equation}
where $R$ is a measure of the size of the system and 
$\sigma$ a measure of the system's velocity dispersion.
Here we take the size of a system to be
the effective radius ($R_e$)  from N14, and 
the central velocity dispersion $\sigma_0$ from Table 1 for AIMSS
objects, Table A1 for UCDs from Mieske et al. (2013) and Taylor
et al. (2010), and N14 for all other objects. We note that using 
the central velocity dispersion over some scaled
aperture, or a weighted global value, facilitates comparison with
other pressure-supported systems. 

To derive the dynamical mass one needs to know the virial
coefficient $C$. A variety of approaches and hence values of $C$
have been adopted in the literature, generally in the range 4 $<$
$C$ $<$ 7.5. 
Here we adopt $C$ = 6.5 to be comparable to the dynamical mass
calculations of Mieske et al. (2013). This value corresponds to
a Sersic index of $n$ $\sim$ 2 (Bertin et al. 2002), 
which is a reasonable value for
UCDs (Taylor et al. 2010) 
and for the objects that we focus on in this paper. Also, as
shown later, such a value of $C$ gives a mean dynamical-to-stellar
mass ratio of close to unity for GCs (which are believed to be free
of dark matter).

\section{Effects of Tidal Stripping}

Before presenting our results, we briefly discuss the effects of
tidal stripping on the properties of small galaxies (e.g. Forbes et
al. 2003) as they are
transformed into cEs and UCDs. Such a scenario is thought to be
the dominant pathway for cEs and the more massive UCDs. 

In general,  tidal stripping will leave the central properties of a
galaxy (e.g. velocity dispersion, metallicity, black hole mass)
relatively unchanged. Here we are particularly interested in the
effects of tidal stripping on the central velocity dispersion $\sigma_0$. In
the seminal work by Bender, Burstein \& Faber (1992) they noted
that {\it ...stripping of stars from the outer parts of a galaxy
will leave $\sigma_0$ approximately constant...}.  
Chilingarian et al. (2009) 
simulated the tidal stripping of a 
disk galaxy embedded within a dark matter halo of 
total mass $\sim10^{12}$ M$_{\odot}$ within the
potential of an M87-like galaxy. 
The final velocity dispersion of the remnant is
within 10\% of the original value even after $\sim$75\% of
the mass is stripped.  

As mentioned in the Introduction, Pfeffer \& Baumgardt (2013)
simulated the tidal stripping of dwarf elliptical galaxies with
nuclei on various orbits.
Using a particle mesh code, they tracked changes of the stellar
mass and size with time 
during the interaction but did not measure the velocity
dispersion. The tidal stripping (`threshing') simulations of 
Bekki et al. (2003) did include velocity 
dispersion but they did not provide details of how of it evolved,  
simply noting that the {\it nucleus remains largely unaffected}. 
We expect $\sigma_0$ to be largely unchanged in the stripping process as the nuclei 
of dEs are relatively tightly bound. 

Below we examine the evolution in dynamical-to-stellar mass ratio
with time for two of the tidal stripping simulations of Pfeffer
\& Baumgardt (2013). In particular, we follow the `stripping
tracks' of a high mass (3 $\times$ 10$^9$
M$_{\odot}$) and low mass (7 $\times$ 10$^8$
M$_{\odot}$) 
dE progenitor, i.e. their simulations 39 and
3 respectively (with data kindly supplied by Baumgardt \& Pfeffer
2014, priv. comm.). 
In the absence of a 
central velocity dispersion, we normalise the dynamical-to-stellar
mass ratio to a value of unity
at the start of each simulation (the model progenitors are dark
matter free). At each time step, of 25 Myr,
we recalculate the mass ratio (using the reduced size and stellar mass
information from the simulations) until the end of the simulation
(typically after a few Gyrs have elapsed). The final remnants have $<$1\% of the mass
of the progenitor, i.e. over 99\% of the stellar mass has been lost
due to stripping. 

The key assumptions in this calculation are: 1) that the central
velocity dispersion is unchanged; 2) that the virial coefficient
is unchanged from the progenitor dE to the remnant; 
and 3) that the remnant is in virial equilibrium
and hence equation 1 is valid. The first assumption appears to be
valid to within $\sim$10\% (20\% in dynamical mass), and the second to
within 40\% (Bertin et al. 2002; Agnello et
al. 2014). 

\section{Results and Discussion}

\subsection{Comparison of dynamical and stellar masses}





Recently, Mieske et al. (2013) re-examined the dynamical and
stellar masses of CenA, Virgo and Fornax UCDs with well-measured properties. 
Before contrasting
our results with theirs we first display their sample in Fig. 2. 
Rather than plot the dynamical-to-stellar mass ratio against mass
(e.g. their figure 2), we prefer to plot dynamical vs stellar mass
to avoid the possibility of any trend being dominated by simply
the same quantity being plotted against itself (i.e. instead of
A/B vs A we plot A vs B). 

\begin{figure}
\begin{center}
\includegraphics[scale=0.33,angle=-90]{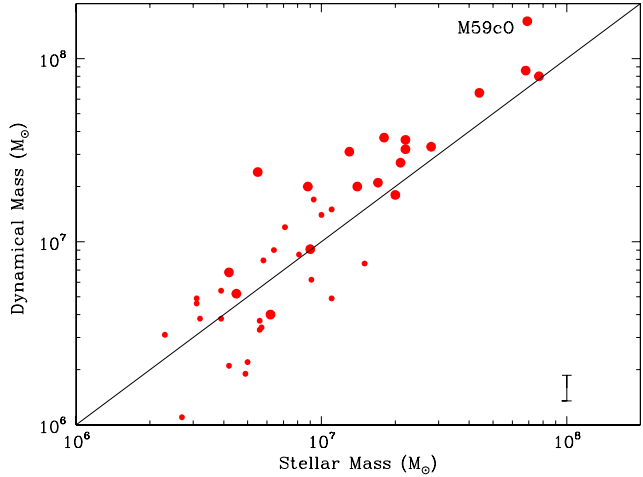}
\caption{Dynamical vs stellar mass for the 
sample of Virgo, Fornax and CenA UCDs using data from Mieske et
al. (2013). Larger circles denote objects with 
R$_e$ $>$ 10 pc. The diagonal solid line
shows a 1:1 relation. A typical observational 
error bar is shown lower right. 
The Mieske et al. sample shows
that below about 10$^7$ M$_{\odot}$ UCDs scatter about the 1:1
relation, i.e. they are consistent with the total (dynamical)
mass being due purely to stars. However, above this mass UCDs
have systematically higher dynamical-to-stellar mass ratios. 
The larger-sized UCDs tend to also be those with the largest stellar
masses. The massive UCD M59cO is labelled. 
}
\end{center}
\end{figure}

Figure 2 uses the dynamical and stellar masses directly from their table
3. Their dynamical mass calculations use a mass model and a profile tailored to each 
object.
They note that in terms of
the formula given by equation 1 above their approach is
equivalent to a light profile with a virial coefficient $C \sim$ 6.5. Their
stellar masses are derived from each UCD's observed M$_V$ and 
[Fe/H] metallicity 
by applying a 13 Gyr, solar alpha abundance, single stellar
population model (actually the mean of Maraston 2005 and
Bruzual \& Charlot 2003) with a Kroupa IMF. 

The plot shows that the data scatter evenly about the unity relation
for masses $<$ 10$^7$ M$_{\odot}$. However, as noted by
Mieske et al., the mass ratio for these objects is bimodal, almost
avoiding ratios of unity. They found the distribution to
be inconsistent with a unimodal distribution at the 99.2\% level
using the KMM test. 
Above 10$^7$ M$_{\odot}$
there is a tendency for the data to lie systematically
above the unity relation (as also seen in figure 2 of Mieske et al. 2013).
Figure 2 
also shows that the high mass sample is dominated by objects with
effective radii
R$_e$ $>$ 10 pc. The location of M59cO is highlighted; the Mieske et al. (2013) values suggest 
it has a large dynamical mass.

\begin{figure}
\begin{center}
\includegraphics[scale=0.33,angle=-90]{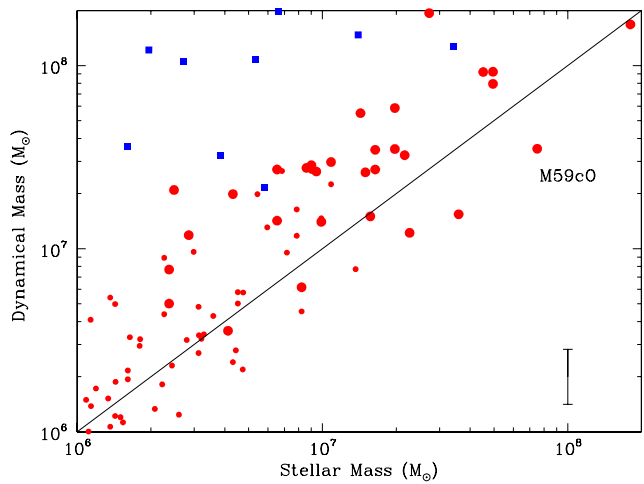}
\caption{Dynamical vs stellar mass for the enlarged
and revised sample of UCDs. Red circles denote UCDs, with larger circles
for objects with R$_e$ $>$ 10 pc. Blue squares denote Local Group
dSph galaxies.  
The diagonal solid line
shows a 1:1 relation. A typical observational 
error bar is shown lower right. 
Excluding the dSph galaxies,
our enlarged sample of UCDs shows the 
trend for elevated mass ratios as 
seen in the smaller Mieske et al. (2013) sample
(Fig. 2), but does not support their claim for a bimodality in the mass ratios. 
The dynamical mass of M59cO is lower compared to the
Mieske et al. sample due to the smaller velocity dispersion
measured by N14. 
}
\end{center}
\end{figure}

In Figure 3 we show our enlarged and revised sample of UCDs (i.e. combining
Table 1 with other data from the literature). 
Dynamical masses are calculated using equation 1 
using the R$_e$ values and central velocity dispersions as listed in Table 1, with
a virial coefficient for all objects assumed to be 
$C$ = 6.5 (this includes the Local Group dSph
galaxies shown in the figure). 
As well as a few UCDs with very high dynamical-to-stellar mass ratios (these are identified and 
discussed further in Section 6.2), the Figure reveals a
few high mass UCDs now scattering below the unity relation. This
includes M59cO with its dynamical mass based on our velocity
dispersion of 25.7 $\pm$ 2.2 km s$^{-1}$. The dynamical mass of
M59cO calculated by Mieske et al. (2013) is much higher as they
used the literature value of 48 $\pm$ 5 km s$^{-1}$ which N14
note was probably incorrect as it is close to the spectral
resolution of the instrument used to obtain it (Chilingarian \&
Mamon 2008). This object
illustrates how the quoted uncertainties in the literature are usually
measurement uncertainties only and do not include systematic errors. An
estimate of the latter can be seen in the scatter of the data
points. With our enlarged and revised sample there is still a tendency for
systematically higher mass ratios for the higher mass UCDs, confirming the 
trend seen in Figure 2. Figure 3 also shows that the transition to higher 
mass ratios for stellar masses of a few 10$^6$ to above 10$^7$ 
M$_{\odot}$ is very much driven by objects with R$_e$ $>$ 10 pc. 
 

\begin{figure}
\begin{center}
\includegraphics[scale=0.33,angle=-90]{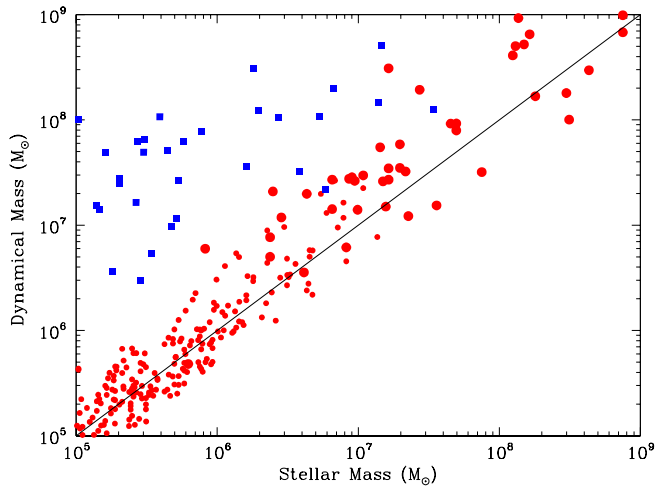}
\caption{Dynamical vs stellar mass for the 
enlarged sample of UCDs and GCs. Red circles denote GCs and UCDs,
with larger circles for 
objects with R$_e$ $>$ 10 pc. 
Blue squares denote Local Group
dSph galaxies.  
The diagonal solid line
shows a 1:1 relation. 
Globular clusters, with stellar masses typically less than 
10$^6$ M$_{\odot}$, scatter evenly about  
the 1:1 relation indicating that their total (dynamical) 
mass is consistent with their stellar mass. 
}
\end{center}
\end{figure}

We do not see any strong evidence for the
bimodal mass ratio trends claimed by Mieske et al. (2013).
To illustrate this further, 
in Figure 4 we show a histogram of the dynamical-to-stellar mass
ratios from Figures 2 and 3. As expected, the Mieske et al. (2013)
data reveal a bimodal mass ratio distribution. Our enlarged sample
does not, revealing a unimodal distribution peaked around unity. 
However, it should be noted that our sample includes objects more 
distant than the CenA, Fornax and Virgo objects of Mieske et al. and 
the resulting uncertainty in the dynamical masses is also greater (as 
indicated by the error bars in Figures 2 and 3). 
So although we do not see any evidence for the bimodality in the mass
ratio of our low-mass UCDs, our uncertainties are higher than those in the 
Mieske et al. study and may have effectively washed-out a weak trend.




Next we extend the parameter space to smaller masses, to include
objects traditionally called globular clusters (GCs), and to
higher masses, but still excluding objects that might be
considered compact ellipticals (i.e. with stellar masses $>$
10$^9$ M$_{\odot}$). Globular
clusters are thought to be free of dark matter (and to lack
massive central black holes) and so should have mass ratios that
scatter about the unity relation. Using GC data for the Milky Way
and M31 from N14, Fig. 5 shows that this is indeed the case. The
trend, and level of scatter, are fairly similar from 10$^5$ to
about 10$^6$ M$_{\odot}$. 
The level of scatter for the GCs ($\le$0.3 dex) gives an indication
of the systematic errors in estimating dynamical-to-stellar mass
ratios in this mass range. Systematic trends for M31 GC mass ratios with
metallicity, which contribute to this scatter, are discussed in
Strader et al. (2011).


Figure 6 extends the mass regime to include elliptical galaxies
(i.e. cEs, dEs and gEs). A few caveats need to be kept in
mind when interpreting this plot: 1) 
as we move from UCDs with Sersic n $\sim$ 2 to gEs with n $\sim$
4, the virial coefficient should be reduced by $\sim$40\% to
account for this change in structure, thus
gradually reducing the dynamical mass (a downward change in
Fig. 6 of $\sim$0.15 dex); 2) the IMF may become
more bottom-heavy in highest mass giant ellipticals (Conroy \& van Dokkum 2012), which
would increase stellar masses for a given luminosity (a rightward
change in Fig. 6); 3) the samples of high mass UCDs and cEs 
are still incomplete (it is not clear if this
incompleteness would affect the trends seen in Figure 6 or
earlier Figures).  

\begin{figure}
\begin{center}
\includegraphics[scale=0.33,angle=-90]{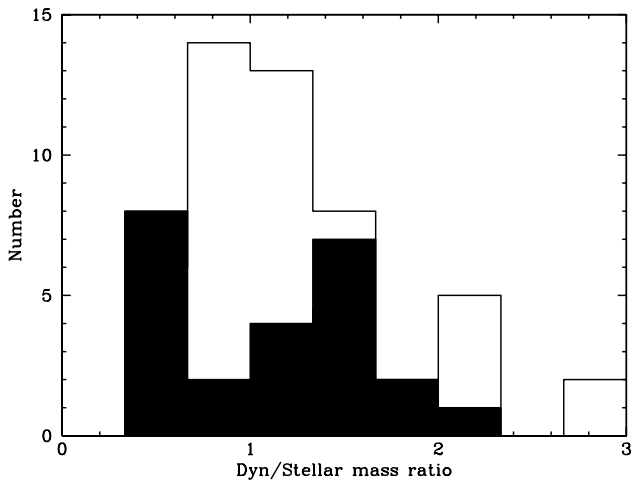}
\caption{Histogram of dynamical-to-stellar mass ratio for
low-mass ($<$ 10$^7$ M$_{\odot}$) UCDs. The
Mieske et al. (2013) sample (from Figure 2), shown as a filled
histogram, shows a bimodal nature to the mass ratios. However,
our enlarged sample (from Figure 3), shown by an open histogram, does not.
}
\end{center}
\end{figure}

In Fig. 6 we see that our sample reaches relatively high masses of
10$^{10}$ M$_{\odot}$ but remains fairly evenly scattered about the
unity relation at the highest masses.  The two cE galaxies with masses
above 10$^{10}$ M$_{\odot}$ come from Chilingarian et al. (2009). The
figure also highlights the need to measure internal kinematics for 
more objects with stellar masses
$\sim$ 10$^8$ M$_{\odot}$ 
(e.g. Forbes et al. 2011). 


\begin{figure}
\begin{center}
\includegraphics[scale=0.33,angle=-90]{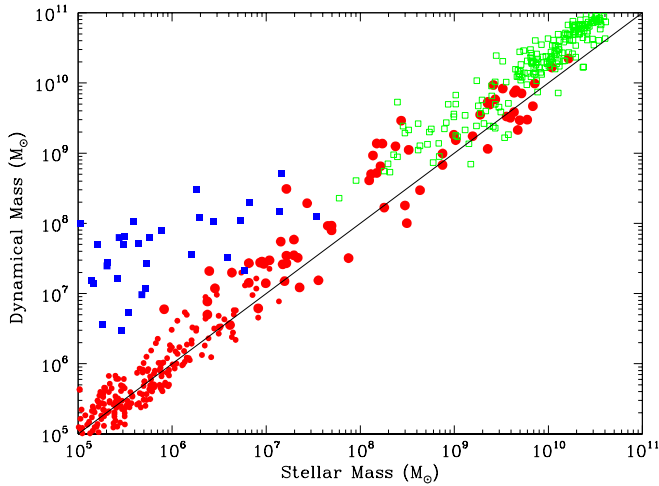}
\caption{Dynamical vs stellar mass for the 
enlarged sample of GCs, UCDs, cEs and galaxies.
Red circles denote GCs, UCDs and cEs, 
with larger circles for 
objects with R$_e$ $>$ 10 pc. 
Blue squares denote Local Group dSph galaxies, and green open squares
denote dE and gE galaxies.   
The diagonal solid line shows a 1:1 relation.
Although most UCDs/cEs with stellar masses around a few 10$^8$ 
M$_{\odot}$ show systematically elevated mass ratios, the ratio
returns to near unity for the most massive UCDs/cEs. 
}
\end{center}
\end{figure}

\subsection{Trends with Stellar Mass}

To better visualise how the average mass ratio changes from GCs
to cEs and to highlight any objects with extreme mass ratios, 
in Figure 7 we plot the mean
dynamical-to-stellar mass ratio as a function of stellar mass. 
We include both AIMSS and literature samples. 
The plot shows half a dozen objects with extremely high mass ratios of 6--11, and a general 
locus of objects with lower mass ratios. First we discuss trends for 
the majority of objects and then focus on the extreme mass ratio objects. 

The GCs have mass ratios close to unity as expected. For UCDs
there is a consistent trend for increasing ratios up to a few 10$^8$
M$_{\odot}$ (albeit with large scatter). We also note a hint of a separate 
sequence of mass ratios around 3--4  
in the stellar mass range $7 \times 10^{5} <$  M$_{\odot}$ 
$< 7 \times 10^{6}$ for a dozen UCDs 
but this may be due simply to small number statistics. 
At the masses associated
with cE galaxies ($>$ 10$^9$
M$_{\odot}$), the mass ratio returns to being close to 
unity. 
Values for the mean mass ratios and the error on the mean,
along with the median values,  
are given in Table 2. The median values, which are less effected by outliers, 
indicate a similar trend to the 
mean values. 

\begin{figure*}
\begin{center}
\includegraphics[scale=0.65,angle=-90]{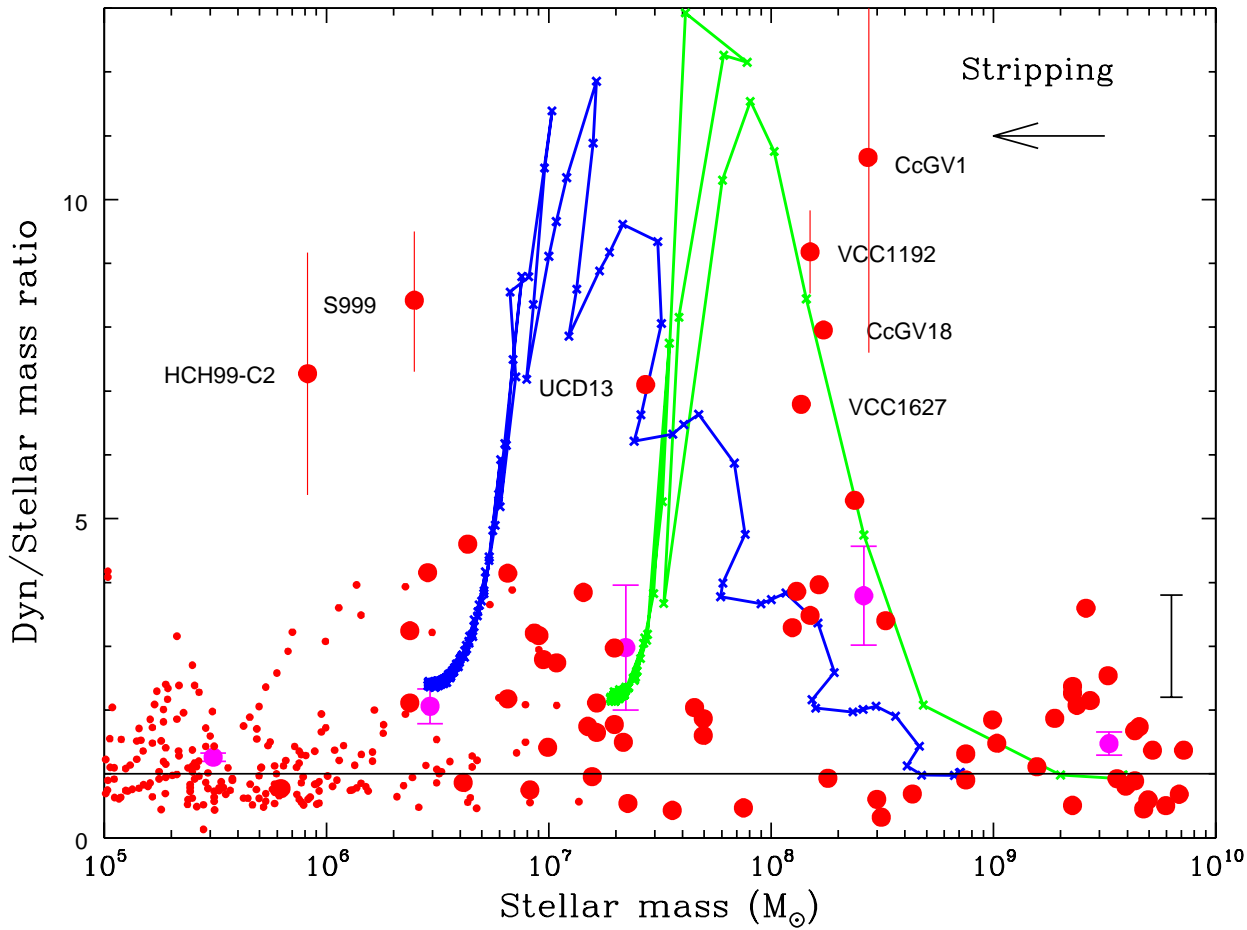}
\caption{Dynamical-to-stellar mass ratio variation with stellar
mass. Red circles show individual GCs, UCDs and cEs, with large circles 
for objects with R$_e$ $>$ 10 pc.  
Large magenta
circles show the mean mass ratio (in bins of 1 dex) with errors on the mean.
The solid black line shows the 1:1 relation of the mass ratio.
The blue and green tracks show the effects of tidal stripping 
on the mass ratio in
25 Myr steps for two dE progenitors of different initial mass 
(the tracks have been normalised to a mass ratio of unity at the start of the
simulation, with $\sigma_0$ assumed to be constant) from the
models of Pfeffer \& Baumgardt (2013). The arrow shows the
general direction of stripping. Several high mass ratio objects, 
HCH99-C2, S999, CcGV1, CcGV18, VCC1192, VCC1627 and Perseus-UCD13 
are labelled (with measurement error 
bars given for the four most extreme objects). A typical error bar for the whole 
sample is shown in the lower right. 
The data show a steady increase in the mean dynamical-to-stellar
mass ratio from GCs
(mass $<10^6$ M$_{\odot}$) to those objects with stellar mass 
$\le$ $10^9$ M$_{\odot}$, and then
a decrease back to around unity for the highest mass cEs. 
The stripping tracks show that as stellar mass is lost from the
progenitor dE galaxies, 
the inferred mass ratio increases in the first
few hundred Myr and then decreases to a value similar to those
of $10^7$ M$_{\odot}$ UCDs.
}
\end{center}
\end{figure*}

Thus we do see some evidence of an upturn in the mass ratio 
above the mass limit of 7 $\times$
10$^7$ M$_{\odot}$ as advocated by N14 
at the transition to objects which are all formerly
stripped galaxies. However, this ratio returns to unity for
masses above 10$^9$ M$_{\odot}$, suggesting that the total mass of
cE galaxies can be accounted for by their stars alone. 
We have estimated the effect of a changing virial coefficient 
for the mass range shown in Fig. 7 and find it is only a $\le$40
per cent effect, and therefore it cannot explain the observed trend. 
It is interesting that the mass ratio for cE galaxies is near unity, as cEs are
generally thought to be the remnant of a larger galaxy (Faber
1973; Chilingarian et al. 2009; 
Huxor et al. 2011). Support for this view was also found
by N14, in which we noted that most cEs have $\sigma_0$ $\sim$ 100
km s$^{-1}$, typical for that of low mass ellipticals (whereas
UCDs have $\sigma_0$ $\sim$ 45
km s$^{-1}$, more typical of dwarf ellipticals).

\begin{table}
\caption{Mean dynamical-to-stellar mass ratios.}
\begin{tabular}{lccl}
\hline
log M$_{\ast}$ & Mean & Median & Object\\ 
(M$_{\odot})$ & Ratio & Ratio & Class\\
\hline
5--6 & 1.26 $\pm$ 0.07 & 1.04 & GC\\
6--7 & 2.06 $\pm$ 0.27 & 1.33 & UCD\\
7--8 & 2.98 $\pm$ 0.98 & 1.78 & UCD\\
8--9 & 3.79 $\pm$ 0.78 & 3.40 & UCD\\ 
9-10 & 1.48 $\pm$ 0.18 & 1.37 & cE\\
\hline

\end{tabular}
\end{table}

As with our previous figures, the larger circles in Figure 7 indicate 
objects with R$_e$ $>$ 10 pc. As noted in the discussion of 
Figure 3, the transition to higher 
mass ratios as stellar mass increases from a few 10$^6$ to above 10$^7$ 
M$_{\odot}$ is very much driven by objects with R$_e$ $>$ 10 pc. 
All of the objects with apparent mass ratios above 5 have sizes R$_e$ $>$ 
10 pc. We have investigated whether the mass ratio shows a continuous 
trend with the measured size for UCDs and cEs, and find none. 
 
Figure 7 also shows the evolutionary `stripping tracks' for two
model galaxies from Pfeffer \& Baumgardt (2013) as described in
Section 5. The tracks have been normalised to a unity mass
ratio at their initial (progenitor) mass. The stripping tracks
show that as stellar mass is lost, the mass ratio first increases
and then declines after several
Gyr. The reason for this behaviour, is that instead of stellar
mass and radius being reduced in lock-step, the simulations show
that the initial stripping preferentially removes mass with
little change in radius, while in the later stages the remnant
becomes much smaller with limited mass loss.  

It is likely that
the extreme mass ratios inferred in the models at early stages in the
stripping, and similar ratios calculated for some UCDs, indicate that
the objects are out of virial equilibrium. In this sense, the mass ratio
for both the simulation tracks and the observed objects should be
considered to be an `apparent' mass ratio and not a physical one.
This is a short-lived
early phase in the stripping process -- objects in this phase should be
relatively rare and may reveal tidal features or extended halos 
in deep imaging. For example, VUCD7 is known to have a dual, core plus
halo, structure. From a single Sersic fit its size is measured to
be R$_e$ $\sim$ 100 pc (Evstigneeva et al. 2007) and we calculate
an apparent mass ratio of $\sim$19 (not shown in Figure
7). However, if we were to use the core size of R$_e$ $\sim$ 10
pc and the stellar mass of the core, then its mass ratio would
reduce to around 6. We note that our stellar population fit to the colours of 
VUCD7 are poor and therefore the dynamical-to-stellar mass ratio is uncertain.
UCD13 in the Perseus cluster is another
example of an object with a core plus halo surface brightness
profile (Penny et al. 2014) and a relatively high inferred mass
ratio from a single component fit (i.e. mass ratio of 7). 
So in some cases, the high apparent mass ratios of UCDs are due to a 
dual-component structure.

A few of the objects with extreme apparent mass ratios are labelled
in Figure 7. They include the CenA group UCD HCH99-C2 (Taylor et
al. 2010), Virgo cluster UCD S999 (Hasegan et al. 2005), the Coma
cluster cEs CcGV1 and CcGV18 (Price et al. 2009), Virgo cluster cEs 
VCC1192 and VCC1627 (Smith Castelli et al. 2013) and Perseus-UCD13 (Penny 
et al. 2014, and this work). 
Thus the measurements come from different 
literature sources and different instruments. 
For the four most extreme objects we show error 
bars which include the measurement uncertainty on R$_e$ and $\sigma_0$, and 
we assume a magnitude uncertainty of $\pm$0.1 mag. The Figure shows that 
even taking into account  
measurement uncertainties these extreme objects have apparently elevated mass ratios.
We have re-checked the transformation into stellar mass for these objects and find 
them to be reasonable.  It is possible that some of these literature objects have an 
overestimated velocity dispersion, as was the case for M59cO, perhaps due to insufficient 
spectral resolution. 

In summary, some of the extreme mass ratios for observed UCDs are due
to a dual-component structure for which a single component fit gives
rise to an inflated size measurement (e.g. VUCD7, Perseus-UCD13).
Some objects may have spurious velocity dispersions in the literature
(e.g. M59c0). For the remainder, it is unclear if measurements are in
error or if the extreme apparent mass ratio is real. 
Deep imaging of these objects for extra-tidal features would be
worthwhile.

It is worth emphasising that the apparent high mass ratios of the simulations 
only occur in the first Gyr. Thus it will be rare to catch a 
galaxy in this early stage of the stripping process. 

At the final stages of the stripping process, after a few Gyr, the
models have mass ratios and stellar masses consistent with 10$^7$
M$_{\odot}$ UCDs. We note that the mass ratio at the end of the
simulation is $\sim$2, as this simply reflects the the size-to-stellar
mass ratio of the remnant to the progenitor. As the models are dark
matter free, the final mass ratio should be unity. Thus our
assumptions of a constant central velocity dispersion and/or virial
coefficient may be incorrect. There may also be some resolution
effects in the model itself, as at times during the stripping process
the galaxy appears to briefly gain stellar mass.  Nevertheless, the
qualitative agreement between the tracks and the distribution of UCDs
suggests that tidal stripping may be an alternative explanation for
the elevated mass ratios and deserves further investigation.



\subsection{Trends with Stellar Surface Density}
 
Next we explore how the dynamical-to-stellar mass ratio varies
with stellar surface density and 
what this may tell us about the
cause of the rising mass ratio with object mass for UCDs. 
We calculate stellar
surface density following N14, i.e. the 
stellar mass divided by the effective surface 
area of 2$\pi$R$_e^2$. 




\begin{figure*}
\begin{center}
\includegraphics[scale=0.65,angle=-90]{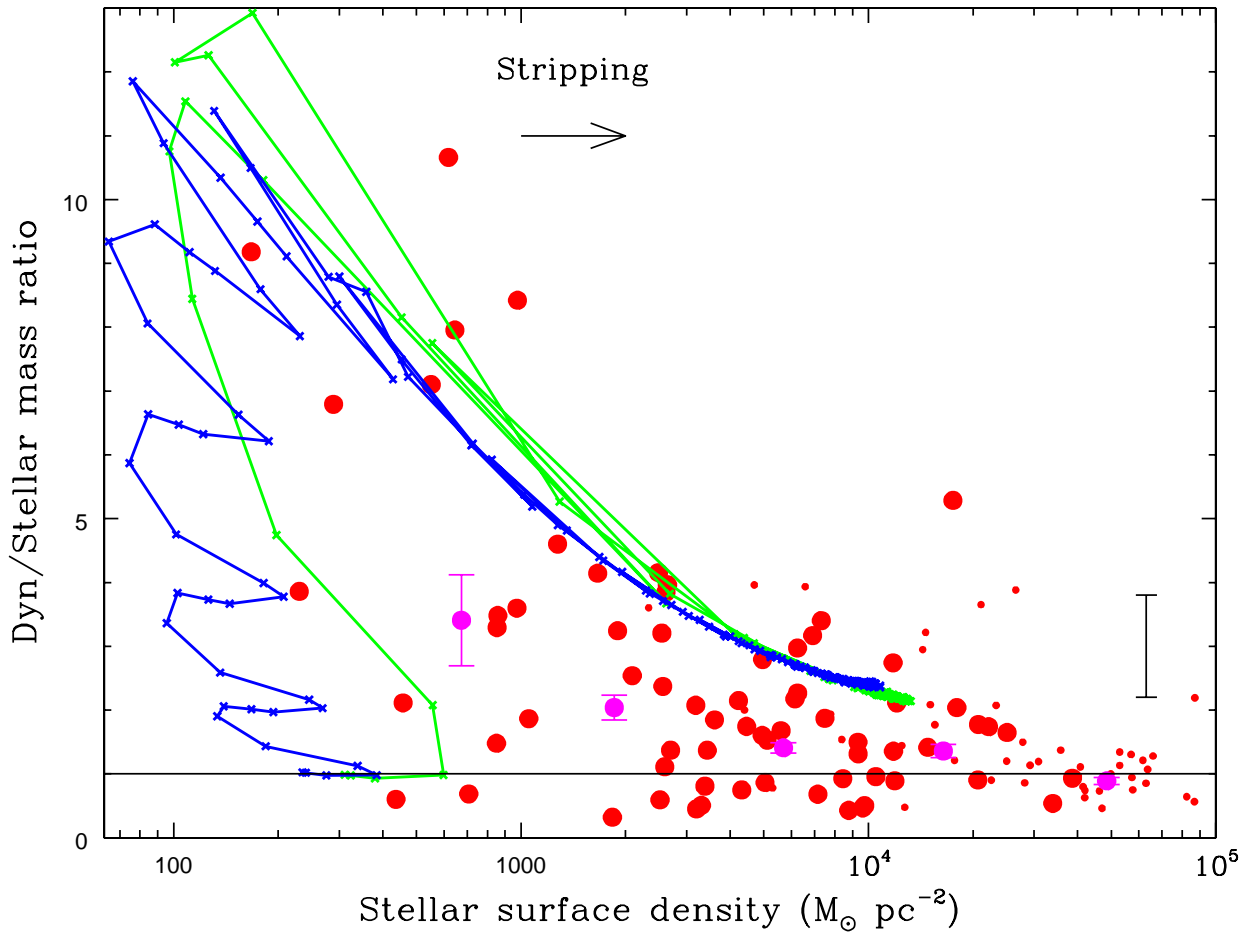}
\caption{Dynamical-to-stellar mass ratio vs stellar surface density.
Red circles show individual UCDs and cEs only, with larger circles for 
objects with R$_e$ $>$ 10 pc. 
Large magenta
circles show the mean mass ratio (in bins of 0.5 dex) with errors on the mean.
The solid black line shows the 1:1 relation of the mass ratio.
A typical error bar for the whole sample is shown in the lower right.
There is a general trend for decreasing mass ratios in the  
higher density systems. 
The blue and green tracks show the effects of tidal stripping 
on the mass ratio in
25 Myr steps for two dE progenitors 
(the tracks have been normalised to a mass ratio of unity at the start of the
simulation, with $\sigma_0$ assumed to be constant) from the
models of Pfeffer \& Baumgardt (2013). The arrow shows the
general direction of stripping. The stripping tracks show
that the progenitors start with a relatively low average density but 
after a few hundred Myr the remnants become more dense as the
nuclei increasingly dominate and the mass ratios get smaller 
as the stripping progresses.}
\end{center}
\end{figure*}

In Figure 8 we show the mass ratio as a function
of stellar surface density for UCDs and cEs only (i.e. we 
exclude GCs and other galaxy types). The plot includes the mean value 
and error on the mean (the median values, not shown, reveal a similar trend). 
The general trend is for the highest density objects to have
the lowest mass ratio, with the very highest density objects
having mean mass ratios of unity. The half dozen UCDs and cEs with extreme mass ratios, and relatively 
low surface densities, are the same as labelled in Figure 7.

\subsubsection{Dark Matter}

The trend in Figure 8 is in the direction expected if
an increased fraction of dark matter is responsible for the
elevated ratios in UCDs. 
Although the density of baryons is so high in the
highest stellar density UCDs that dark matter cannot be
accommodated (Mieske et al. 2008), it may make a progressively larger contribution in
the more diffuse objects. Detailed modelling is required to confirm this possibility. 

\subsubsection{Central Black Holes}

According to Mieske et al. (2013), the elevated mass ratios for UCDs
could be explained by central black holes that contribute 10-15 per
cent of UCD masses. Most recently, strong evidence for a central black
hole in a UCD has been reported from X-ray emission (Strader et
al. 2013) and central velocities obtained by an integral field unit
(IFU) with adaptive optics (Seth et al. 2014). The black hole
represents 15 per cent of the mass of the UCD. Interestingly this
object, named M60-UCD1, has a very high central density ($\sim10^5$
M$_{\odot}$ pc$^{-2}$) but has a mass ratio close to unity (see Table
1). In this case, the velocity dispersion integrated within an
aperture does not reveal the high velocities associated with a central
black hole. It is also worth mentioning UCD3 in the Fornax cluster
which has been observed with an IFU by Frank et al. (2011). They found
no evidence for a black hole to within 5 per cent of the UCD mass.
Clearly more UCDs require follow-up with the resolution sufficient to
resolve the sphere of influence of the black hole.

Central black holes are the norm in high mass galaxies. If the trend
seen in Figures 7 and 8 is due to the presence of black holes, then
they must be preferrentially dominating the dynamical mass (via the
observed velocity dispersion) of the higher mass and lower density
UCDs respectively. Larger relative contributions from the black hole
would be expected in the systems that have been stripped of the most
stars. It isn't obvious why this would be the case for the higher mass
{\it and} lower density UCDs.  Strangely, the cE galaxies, which are thought
to come from higher mass progenitors than UCDs (and hence possess
black holes), do not reveal elevated mass ratios.

\subsubsection{Non-universal IMF}

A non-universal IMF may give rise to the elevated mass ratios seen for UCDs. 
The IMF is observed to be more bottom-heavy in the
cores of giant elliptical galaxies, revealing a strong trend with
central velocity dispersion for $\sigma_0$ $\ge$ 200 km
s$^{-1}$ (La Barbera et al. 2013; Ferreras et al. 2013; Conroy
et al. 2013). The most massive ellipticals tend to have relatively low 
density cores, compared to lower mass ellipticals (Graham \& Guzman 1993).  
Thus we might expect a
trend for more bottom-heavy stellar populations 
in galaxies with lower central densities.
A trend of this nature is seen in Figure 8, 
with UCDs and cEs having higher mass
ratios at lower surface densities.
However, all of the UCDs and cEs
have $\sigma_0$ $<$ 200 km s$^{-1}$ and are therefore
unlikely to be the remnant cores of giant ellipticals. 

In the case of a top-heavy IMF, Dabringhausen 
et al. (2010) showed that high central densities (which promoted encounters between proto-stars) 
were associated with a top-heavy IMF. Assuming that the current stellar densities of UCDs 
reflect their initial central densities, then the trend seen in Figure 8 is the opposite of that  
expected for a top-heavy IMF.

\subsubsection{Tidal Stripping}

In Section 6.2 we suggested that tidal stripping was another
possible reason for the elevated mass ratios of UCDs.  The
`stripping tracks' of the same two model galaxies from Figure 7
are shown in Figure 8. The model galaxies start out at relatively
low stellar density, they quickly rise in apparent mass ratio with little
change in stellar density. The simulations may not be in virial
equilibrium during these early stages of rapid change. This may also 
be true for the objects with apparently extreme mass ratios (see Figure 7 
for the identity of several such objects). 

In the later stages of the stripping process the simulations 
decline in mass ratio
and increase in density, with both models ending up with a mass
ratio and density similar to that of the average UCD.
This latter stage evolution is qualitatively 
similar to the trend seen in the mass ratio of UCDs and cEs. \\



In summary, the trend for higher dynamical-to-stellar mass ratios in lower surface density UCDs and cEs is 
qualitatively consistent with an increased contribution of dark matter, stellar populations with a 
top-heavy IMF and tidal stripping. It is less likely to be explained by a bottom-heavy IMF or an 
increasing contribution of a central black hole (despite good evidence for a black hole in at least 
one UCD).

\section{Conclusions and Future Work}

Here we present central velocity dispersions and dynamical
masses for a sample of 27 UCDs and cEs recently discovered by 
the AIMSS survey. These data are combined with literature data to
provide the largest sample of UCDs and cEs with internal kinematic
measurements and dynamical masses. Such objects have properties that
are intermediate between traditional star clusters and small
galaxies. 
Using this expanded and revised sample, we
re-examine the elevated dynamical-to-stellar mass ratios in
UCDs.

Although the
scatter is large, we confirm that the dynamical-to-stellar 
mass ratios of UCDs are
systematically higher than for globular clusters. 
However, at the very highest masses in our sample,
i.e. $\ge$ 10$^9$ M$_{\odot}$, we find that cE galaxies have mass
ratios close to unity consistent with being composed largely of 
stars. 
We also re-examine claims for a 
bimodal mass ratios among low-mass (i.e. 
less than 10$^7$ M$_{\odot}$) UCDs. Although we find no evidence to support
this in our combined sample, 
our larger uncertainties may have `washed-out' a weak trend. 

In the literature various possible reasons for the elevated mass
ratios in UCDs have been put forward. These include a variable
IMF, a central massive black hole and the presence of dark
matter. 
Here we present another possible reason for the
elevated mass ratios, i.e. tidal stripping.

We find that the {\it final} 
stellar mass, stellar density and dynamical-to-stellar 
mass ratio of the tidally stripped dE progenitors in the  
simulations of Pfeffer \& Baumgardt (2013) 
are in reasonable agreement with the typical 
observed values for UCDs (under the 
assumption that the central
velocity dispersion and virial coefficient are largely unchanged in the 
process). 
However, at early stages in the stripping process, the galaxy is
probably not in virial equilibrium. This may explain the apparent 
extreme dynamical-to-stellar mass ratios of up to 10 for some
objects. We also note that some of the extreme mass ratio objects reveal 
dual-component structures and/or may have spurious measurements in the 
literature.

Despite some tantalizing results, there is a clear need for
detailed realistic modelling of tidal
stripping. Models need to be extended in mass
to include higher mass progenitors (the likely progenitors of cE
galaxies). As the
morphology of the progenitor has an important effect on the
efficiency of stripping and the structure of the remnant,  
a range of central densities needs to be explored. 
A range of orbits from circular to plunging radial orbits, 
within different gravitational potentials, should be modelled. 
The properties of the remnant that need to be predicted, in
addition to size and luminosity (Pfeffer \& Baumgardt 2013),
include age, metallicity, colour, dark matter mass, and
particularly the internal kinematics (e.g. central velocity
dispersion) and the resulting dynamical mass. 

On the observational side, confirming the presence of dark matter
in UCDs will be very challenging. However, exploring the other
possible explanations for the elevated mass ratios hold more
promise. IMF variations in UCDs can be constrained by obtaining
very high signal-to-noise spectra that include IMF sensitive
features such as the NaI and Wing Ford bands at red
wavelengths (van Dokkum \& Conroy 2010). Adaptive optics 
observations of the cores of UCDs
can be used to obtain the central velocity fields and hence
search for the rapid motions associated with a massive black
hole (Frank et al. 2011; Seth et al. 2014). 
If the stripping scenario is correct, we expect UCDs in the
early stages of stripping to reveal the presence of extra-tidal
features, such as halo structures or tidal tails. 

\section{Acknowledgements}

We thank J. Pfeffer and 
H. Baumgardt for supplying data from their tidal stripping
simulations, and H. Baumgardt and S. Mieske for supplying their data on UCDs.
We thank J. Janz for useful comments. 
We thank the staff of the W. M. Keck Observatory for their
support. Some the data presented herein were obtained at the
W.M. Keck Observatory, which is operated as a scientific
partnership among the California Institute of Technology, the
University of California and the National Aeronautics and Space
Administration. Based on observations obtained at the Southern
Astrophysical Research (SOAR) telescope, which is a joint project
of the Minist\'{e}rio da Ci\^{e}ncia, Tecnologia, e
Inova\c{c}\~{a}o (MCTI) da Rep\'{u}blica Federativa do Brasil,
the U.S. National Optical Astronomy Observatory (NOAO), the
University of North Carolina at Chapel Hill (UNC), and Michigan
State University (MSU).
Some of the observations reported in this paper were obtained
with the Southern African Large Telescope (SALT).
This paper makes use of data obtained as part of Gemini
Observatory programs GS-2004A-Q-9 and GS-2011A-Q-13.
Based on observations obtained at the Gemini Observatory, which
is operated by the Association of Universities for Research in
Astronomy, Inc., under a cooperative agreement with the NSF on
behalf of the Gemini partnership: the National Science Foundation
(United States), the National Research Council (Canada), CONICYT
(Chile), the Australian Research Council (Australia),
Minist\'{e}rio da Ci\^{e}ncia, Tecnologia e Inova\c{c}\~{a}o
(Brazil) and Ministerio de Ciencia, Tecnolog\'{i}a e
Innovaci\'{o}n Productiva (Argentina).
This work is partly based on observations made with the Isaac
Newton Telescope (INT) operated on the island of La Palma by the
Isaac Newton Group (ING) in the Spanish Observatorio del Roque de
los Muchachos of the Instituto de Astrofisica de Canarias (IAC).
DAF thanks the ARC for financial support via DP130100388. This
work was supported by National Science Foundation grant AST-1109878.

\section{References}

Agnello A., Evans N.~W., Romanowsky A.~J., 2014, arXiv, arXiv:1401.4462 \\
Baumgardt, H., Miseke, S., 2008, MNRAS, 391, 942\\
Bekki K., Couch W.~J., Drinkwater M.~J., Shioya Y., 2003, MNRAS, 344, 399 \\
Bender R., Burstein D., Faber S.~M., 1992, ApJ, 399, 462 \\
Bertin G., Ciotti L., Del Principe M., 2002, A\&A, 386, 149\\
Brodie J.~P., Romanowsky A.~J., Strader J., Forbes D.~A., 2011, AJ, 142, 
199 \\
Bruzual G., Charlot S., 2003, MNRAS, 344, 1000 \\
Br{\"u}ns R.~C., Kroupa P., 2012, A\&A, 547, A65\\
Cappellari M., et al., 2011, MNRAS, 413, 813 \\
Chiboucas K., et al., 2011, ApJ, 737, 86 \\
Chilingarian I., Cayatte V., Chemin L., Durret F., Lagan{\'a} T.~F., Adami C., Slezak E., 2007, A\&A, 466, L21\\
Chilingarian I.~V., Mamon G.~A., 2008, MNRAS, 385, L83\\
Chilingarian I., Cayatte V., Revaz Y., 
Dodonov S., Durand D., Durret F., Micol A., Slezak E., 2009, Sci,
326, 1379 \\
Conroy C., van Dokkum P.~G., 2012, ApJ, 760, 71 \\
Conroy C., Dutton A.~A., Graves G.~J., Mendel J.~T., van Dokkum P.~G., 
2013, ApJ, 776, L26 \\
Dabringhausen J., Hilker M., Kroupa P., 2008, MNRAS, 386, 864 \\
Dabringhausen J., Kroupa P., Pflamm-Altenburg, J., Mieske, S., 2012, ApJ, 747, 72\\
Evstigneeva E.~A., Gregg M.~D., Drinkwater 
M.~J., Hilker M., 2007, AJ, 133, 1722 \\
Faber S.~M., 1973, ApJ, 179, 423\\
Ferreras I., La Barbera F., de la Rosa 
I.~G., Vazdekis A., de Carvalho R.~R., Falc{\'o}n-Barroso J., Ricciardelli 
E., 2013, MNRAS, 429, L15 \\
Forbes D.~A., Beasley M.~A., Bekki K., Brodie J.~P., Strader J., 2003, Sci, 
301, 1217 \\
Forbes D.~A., Lasky P., Graham A.~W., Spitler L., 2008, MNRAS,
389, 1924 \\
Forbes D.~A., Lasky P., Graham A.~W., Spitler L., 2011, MNRAS, 415, 2976 \\
Forbes D.~A., Pota V., Usher C., Strader J., Romanowsky A.~J., Brodie 
J.~P., Arnold J.~A., Spitler L.~R., 2013, MNRAS, 435, L6 \\
Frank M.~J., Hilker M., Mieske S., Baumgardt H., Grebel E.~K., Infante L., 
2011, MNRAS, 414, L70 \\
Geha M., Guhathakurta P., van der Marel R.~P., 2002, AJ, 124,
3073 \\
Geha M., Guhathakurta P., van der Marel R.~P., 2003, AJ, 126, 1794\\
Goerdt T., Moore B., Kazantzidis S., Kaufmann T., Macci{\`o} A.~V., Stadel 
J., 2008, MNRAS, 385, 2136 \\
Graham A.~W., Guzm{\'a}n R., 2003, AJ, 125, 2936 \\
Harris W.~E., 2010, arXiv:1012.3224\\
Ha{\c s}egan M., et al., 2005, ApJ, 627, 203 \\
Kannappan S.~J., et al., 2013, ApJ, 777, 42 \\
La Barbera F., Ferreras I., Vazdekis A., 
de la Rosa I.~G., de Carvalho R.~R., Trevisan M., Falc{\'o}n-Barroso J., 
Ricciardelli E., 2013, MNRAS, 433, 3017 \\
Maraston C., 2005, MNRAS, 362, 799 \\
Mieske S., Dabringhausen, J., Kroupa P., Hilker, M., Baumgardt, H., 2008, AN, 329, 964\\
Mieske S., et al., 2008, A\&A, 487, 921 \\
Mieske S., Frank M.~J., Baumgardt H., L{\"u}tzgendorf N.,
Neumayer N., Hilker M., 2013, A\&A, 558, A14 \\
Misgeld I., Hilker M., 2011, MNRAS, 414, 3699 \\
Murray N., 2009, ApJ, 691, 946 \\
Navarro J.~F., Frenk C.~S., White S.~D.~M., 1997, ApJ, 490, 493 \\
Norris M.~A., Kannappan S.~J., 2011, MNRAS, 414, 739\\
Norris, M., et al. 2014, MNRAS, submitted\\
Penny S.~J., Forbes D.~A., Strader J., Usher C., Brodie J.~P., Romanowsky 
A.~J., 2014, MNRAS, 439, 3808 \\
Pfeffer J., Baumgardt H., 2013, MNRAS, 433, 1997 \\
Phillipps S., Young A.~J., Drinkwater 
M.~J., Gregg M.~D., Karick A., 2013, MNRAS, 433, 1444 \\
Price J., et al., 2009, MNRAS, 397, 1816 \\
Seth, A., et al. 2014, Nature, submitted\
Smith Castelli A.~V., Gonz{\'a}lez N.~M., 
Faifer F.~R., Forte J.~C., 2013, ApJ, 772, 68 \\
Strader J., Caldwell N., Seth A.~C., 2011, AJ, 142, 8 \\
Strader M.~J., et al., 2013, ApJ, 775, L6 \\
Taylor M.~A., Puzia T.~H., Harris G.~L., Harris W.~E., Kissler-Patig M., 
Hilker M., 2010, ApJ, 712, 1191 \\
Tollerud E.~J., Bullock J.~S., Graves 
G.~J., Wolf J., 2011, ApJ, 726, 108 \\
Tollerud E.~J., Geha M.~C., Vargas L.~C., 
Bullock J.~S., 2013, ApJ, 768, 50 \\
Toloba E., Boselli A., Peletier R.~F., Falc{\'o}n-Barroso J., 
van de Ven G., Gorgas J., 2012, A\&A, 548, A78 \\
van Dokkum P.~G., Conroy C., 2010, Natur, 468, 940\\
Willman B., Strader J., 2012, AJ, 144, 76\\

\section{Appendix}

\setcounter{table}{0}
\renewcommand{\thetable}{A\arabic{table}}

\begin{table}
\caption{Central velocity dispersion of ultra compact dwarfs from the literature.}
\begin{tabular}{lc}
\hline
ID & $\sigma_0$ \\ 
 & (km/s) \\
\hline  
                         UCD1             &   41.6$\pm$1.0\\                
                         UCD6             &   28.0$\pm$1.0   \\              
                     UCD2/F-1             &   27.4$\pm$0.6 \\          
                               UCD28/F-5  &   31.9$\pm$1.0 \\               
                                    F-34  &   19.0$\pm$1.5 \\               
                               UCD31/F-6  &   16.3$\pm$2.1 \\               
                                    F-51  &   28.1$\pm$1.6 \\               
                               UCD33/F-7  &   14.4$\pm$1.6 \\               
                               UCD36/F-8  &   34.1$\pm$1.3 \\               
                               UCD39/F-9  &   34.5$\pm$1.6 \\               
                                    F-11  &   28.9$\pm$1.0 \\                
                                    F-17  &   34.1$\pm$1.4 \\               
                                    F-53  &   18.8$\pm$1.8 \\               
                              UCD46/F-22  &   41.9$\pm$1.5	\\	 
                               UCD4/F-24  &   35.1$\pm$1.0 \\               
                                    UCD5  &   26.8$\pm$2.6	\\	      
                                   VUCD1   &   40.3$\pm$1.7 \\  
                                    S999   &   26.2$\pm$1.3 \\  
                                   H8005   &    10.8$\pm$1.9 \\ 
                                    S928   &   23.8$\pm$0.9 \\   
                              VUCD3/S547   &   55.2$\pm$1.5 \\  
                                    S490   &   51.9$\pm$2.7 \\   
                                    S417   &   36.2$\pm$1.5 \\  
                                   VUCD4   &   28.1$\pm$2.0 \\  
                                    S314   &   44.4$\pm$1.4 \\  
                                   VUCD5   &   33.4$\pm$1.6 \\  
                                   VUCD6   &   31.7$\pm$1.8 \\  
                                   VUCD7   &   45.1$\pm$4.1 \\  
                       HGHH92-C29/GC0041   &   20.7$\pm$1.8 \\                
                       HGHH92-C11/GC0077   &   20.3$\pm$1.5 \\  
                                VHH81-C3   &   20.6$\pm$1.1 \\  
                          f2.GC61/GC0150   &   20.6$\pm$1.2 \\  
                                VHH81-C5   &   17.5$\pm$2.2 \\  
                         HCH99-C2/GC0171   &   18.8$\pm$3.0 \\   
                               HGHH92-C6   &   25.7$\pm$1.5 \\  
                        HCH99-C15/GC0213   &   33.4$\pm$5.9 \\   
                        HCH99-C18/GC0225   &   24.1$\pm$1.3 \\   
                       HGHH92-C17/GC0265   &   24.3$\pm$2.9 \\  
                       HGHH92-C21/GC0320   &   22.9$\pm$1.4 \\  
                       HGHH92-C22/GC0326   &   23.3$\pm$1.8 \\  
                       HGHH92-C23/GC0330   &   51.2$\pm$3.7 \\  
                        HGHH92-C7/GC0365   &   28.2$\pm$2.4 \\  

\hline
\end{tabular}
\\
Notes: Object name (ordered by 
Fornax, Virgo and CenA objects), and central
velocity dispersion from Mieske \& Baumgardt (2014, priv. comm.)
and Taylor et al. (2010). 
\end{table}

\end{document}